# Fabrication and Response of High Concentration SIMPLE Superheated Droplet Detectors with Different Liquids

M. Felizardo[1,2], T. Morlat[3], J.G. Marques[4,2], A.R. Ramos[4,2], TA Girard[2,†],

A. C. Fernandes[4,2], A. Kling[4,2], I. Lázaro[2], R.C. Martins[5], J. Puibasset[6]

( for the SIMPLE Collaboration )

[1] Department of Physics, Universidade Nova de Lisboa, 2829-516 Monte da Caparica, Portugal
[2] Centro de Física Nuclear, Universidade de Lisboa, 1649–003 Lisbon, Portugal
[3] Ecole Normale Superieur de Montrouge, 1 Rue Aurice Arnoux, 92120 Montrouge, France
[4] Instituto Tecnológico e Nuclear, IST, Universidade Técnica de Lisboa, EN 10, 2686-953 Sacavém, Portugal
[5] Instituto de Telecomunicações, IST, Av. Rovisco Pais 1, 1049-001 Lisbon, Portugal
[6] CRMD-CNRS and Université d'Orléans, 1 bis Rue de la Férollerie, 45071 Orléans, France

[†] corresponding author: criodets@cii.fc.ul.pt

## Abstract

The combined measurement of dark matter interactions with different superheated liquids has recently been suggested as a cross-correlation technique in identifying WIMP candidates. We describe the fabrication of high concentration superheated droplet detectors based on the light nuclei liquids $C_3F_8$, $C_4F_8$, $C_4F_{10}$ and $CCl_2F_2$, and investigation of their irradiation response with respect to $C_2ClF_5$. The results are discussed in terms of the basic physics of superheated liquid response to particle interactions, as well as the necessary detector qualifications for application in dark matter search investigations. The possibility of heavier nuclei SDDs is explored using the light nuclei results as a basis, with $CF_3I$ provided as an example.

## 1. Introduction

The direct search for weakly interacting massive particle (WIMP) dark matter is generally based on one of five techniques: scintillators, semiconductors, cryogenic bolometers, noble liquids and superheated liquids. The last, in contrast to the others, relies on the stimulated transition of a metastable liquid to its gas phase by particle interaction: because the transition criteria are thermodynamic, the devices can be operated at temperatures and/or pressures at which they are generally sensitive to only fast neutrons, α's and other high linear energy transfer (LET) irradiations.

Only three WIMP search efforts employ the superheated liquid technique: PICASSO [1], COUPP [2] and SIMPLE [3], using $C_4F_{10}$, $CF_3I$ and $C_2ClF_5$ respectively. Of the three, COUPP is based on bubble chamber technology: only PICASSO and SIMPLE employ superheated droplet detectors (SDDs). Because of their fluorine content and fluorine's high proton spin sensitivity, as well as their otherwise light nuclei content relative to Ge, I, Xe, W and others, they have generally contributed most to the search for spin-dependent WIMP-proton interactions. COUPP, with $CF_3I$, has also made a significant impact in the spin-independent sector.

A SDD consists of a uniform dispersion of micrometric-sized superheated liquid droplets homogeneously suspended in a hydrogenated, viscoelastic gel matrix. The phase transition generates a millimetric-sized gas bubble which can be recorded by either optical, acoustic or chemical means; both SDD experiments employ acoustic, while COUPP employs both acoustic and optical (the liquid is essentially transparent, whereas the gel matrix of the SDDs is at best translucent).

The significant difference between the two approaches is that SDDs are continuously sensitive for extended periods since the overall liquid droplet population is maintained in steady-state superheated conditions despite bubble nucleation of some droplets, whereas in the bubble chamber the bulk liquid is only sensitized between nucleation events, each of which precipitates the transition of the liquid volume hence requires recompression to re-establish the metastable state and leads to measurement deadtime. The advantage of the chamber approach is an ability to instrument large active target masses. SDDs have generally been confined to low concentration (< 1 wt% : liquid-to-colloid mass ratio) devices, for use in neutron [4-11], and heavy ion [12] detector applications, with impact in heavy ion and cosmic ray physics, exotic particle detection and imaging in cancer therapy [13,14]. For rare event applications such as a WIMP search, however, higher concentration detectors are required: the PICASSO devices are ~ 1 wt% concentrations. SIMPLE detectors in contrast are generally of 1-2 wt%; concentrations above 2 wt%, in which the droplets are sufficiently close in proximity, tend to self-destruct as a result of massive sympathetic bubble nucleation and induced fractures.

Recently, variation of the target liquids with different sensitivities to the possible scalar and axial vector components of a WIMP interaction has been suggested as a technique in identifying WIMP candidates [15], specifically in the case of COUPP in combined measurements using $CF_3I$ and $C_4F_{10}$. This measurement variation while maintaining equivalent sensitivities in the case of SDDs is not trivial, since device fabrication and operation depends on the individual thermodynamic characteristics of each liquid.

SIMPLE SDD fabrications generally proceed on the basis of density-matching the liquid with a 1.3 g/cm$^3$ food-based gel with low U/Th contamination: a significant difference in gel and liquid densities (as occurs with heavier nuclei liquids) results in inhomogeneous distributions of differential droplet sizes within the detector. Although this has been addressed by SIMPLE via viscosity-matching the gel [16,17], this approach is constrained by the SIMPLE gel melting at 35ºC, limiting the temperature range of the device and hence restricting the liquids employed. The traditional addition of heavy salts such as CsCl to raise the gel density, as originally used by PICASSO with its polyacrylamide-based gels [18], is discouraged since this generally adds radioactive contaminants which must be later removed chemically with the highest efficiency possible.

Thus, the question of liquid variation in SDDs naturally raises the questions of whether or not such "other" SDDs can in fact be fabricated, much less operated, and with what sensitivity. We here describe our fabrications and testing of small volume (150 ml), high concentration (1-2 wt%) SDD prototypes with $C_3F_8$, $C_4F_8$, $C_4F_{10}$, $CCl_2F_2$ and $CF_3I$ including for completeness a "standard" $C_2ClF_5$ device of the SIMPLE dark matter search effort [3]. Section 2 provides an overview of the device fabrication, and describes the experimental testing of the products. The response of superheated liquids to irradiations in general, and liquid characteristics necessary to dark matter searches is discussed in Sec. 3, and applied to the fabricated SDD test results, with the salient aspects of particle discrimination as observed by SIMPLE identified in Sec. 4. Section 5 discusses the considerations necessary to the fabrication and implementation of heavier nuclei SDDs, to include the introduction of a figure of merit based on the light nuclei results by which an initial screening of possibilities can be made in the absence of a complete thermophysical description of the liquids: The fabrication and analysis of a $CF_3I$ is described as an example. Conclusions are formed in Sec. 6.

## 2. Light Nuclei Detectors

For light liquids, SDD construction generally consists of two parts: the gel, and the liquid droplet suspension. The variation of the liquid densities with temperature is shown in Fig. 1, and can be divided into three basic density groups:

(i)      $C_2ClF_5$, $C_3F_8$ ,
(ii)     $CCl_2F_2$ ,
(iii)    $C_4F_{10}$, $C_4F_8$.

For those in groups (i) and (ii) with $\rho \sim 1.3$ g/cm$^3$, small variations in the current $C_2ClF_5$ recipes are indicated; for the more dense liquids of group (iii), viscosity matching is necessary using an additive as discussed in detail in [16,17].

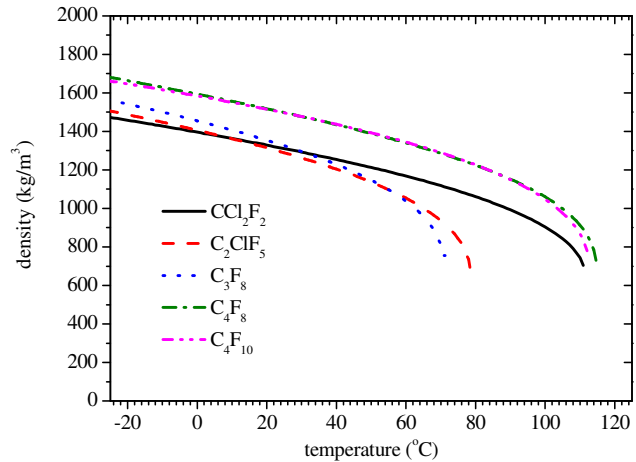

Fig. 1: variation of liquid densities with temperature [19].

### 2.1 Gel Fabrications

The basic SDD ingredients have been described previously [20]. In the density-matched, "standard" case of $C_2ClF_5$, the gel composition is 1.71 wt% gelatin, 4.18 wt% polyvinylpyrrolidone (PVP), 15.48 wt% bi-distilled water and 78.16 wt% glycerin. The gelatin is selected on the basis of its organ origins to minimize the U/Th impurity content; the glycerin serves to enhance the viscosity and strength of the gel, and wet the container surfaces. The presence of the PVP (i) assists in fracture control by viscosity enhancement which decreases diffusion, (ii) improves the SDD homogeneity and reduces the droplet sizes via its surfactant behavior, (iii) decreases the liquid solubility [21], (iv) inhibits clathrate hydrate formation, and (v) reduces the migration of α-emitters to droplet boundaries as a result of actinide complex ion polarity [22].

The basic process, minus several proprietary aspects, has been described in [20]. The ingredients are first formed: powdered gelatin (Sigma Aldrich G-1890 Type A), bi-distilled water and pre-eluted ion exchange resins for actinide removal are combined and left for 12-15 hrs at 45ºC with slow agitation to homogenize the solution. Separately, PVP (Sigma Aldrich PVP-40T) and exchange resins are added to bi-distilled water, and stirred at ~65ºC for 12-15 hrs. Resins and glycerin (Riedel-de-Haën Nº 33224) are combined separately, and left in medium stirring at ~50ºC for 12-15 hrs.

The PVP solution is then slowly added to the gel solution ("concentrated gel"), and slowly agitated at 55-60ºC for 2 hrs. The resins in all are next removed separately by filtering (Whatman 6725-5002A). The glycerin and concentrated gel are then combined at ~60ºC, outgassed at ~ 70ºC, and foam aspirated to eliminate trapped air bubbles. The solution is left at 48ºC for 14 hrs with slow agitation to prevent bubble formation.

For the viscosity-matched protocol required for the $C_4F_8$ and $C_4F_{10}$, the gel composition is essentially the same as in the density-matched recipe, with a small agarose (Sigma Aldrich A0576) addition effected by combining it with glycerin at 90ºC, then adding it to the concentrated gel mix prior its filtration.

Following resin purification, the gel yields measured U/Th contamination levels of < 8.7 mBq/kg $^{238}$U, < 4.9 mBq/kg $^{235}$U and < 6.9 mBq/kg $^{234}$U.

**2.2    Droplet Suspension Fabrications**

The specific protocol for fabrication of a liquid droplet suspension depends on the thermodynamic properties of the liquid. The process with $C_2ClF_5$ is schematically shown in Fig. 2; the temperatures and pressures differ for each liquid.

Following transfer of the gel to the detector bottle, the bottle is first weighed and then removed to a container encased by a copper serpentine for cooling, positioned on a hotplate within a hyperbaric chamber. Once stabilized at 35ºC, the pressure is quickly raised to just above the vapor pressure (~11 bar) of the liquid with continued slow agitation. After thermalization, the agitation is stopped and the liquid injected into the gel through a flowline immersed in ice to simultaneously condense and distill it, and a 0.2 µm microsyringe filter (Gelman Acrodisc CR PTFE 4552T).

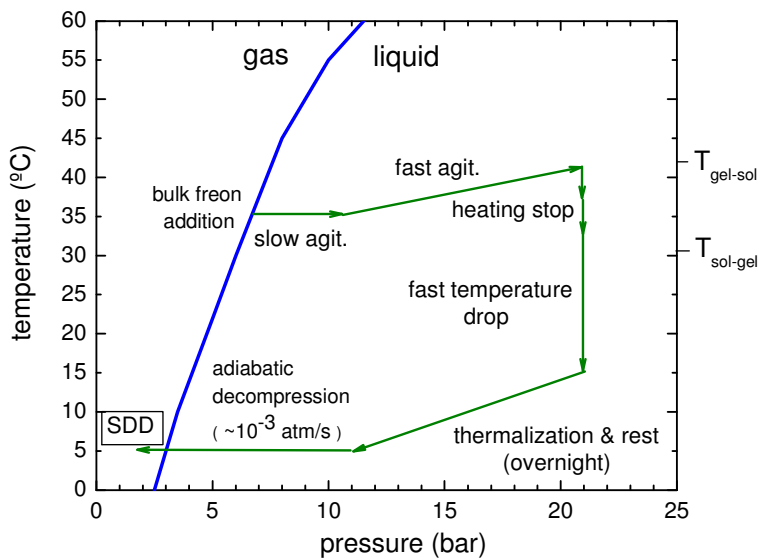

Fig. 2: variation of temperature and pressure following liquid injection in the fabrication of a $C_2ClF_5$ SDD.

Once injected, the pressure is quickly raised to 21 bar to prevent the liquid droplets from rising to the surface, and a rapid agitation simultaneously initiated to shear big droplets; simultaneously, the temperature is raised to 39ºC to create a temperature gradient inside the matrix and to permit dispersion of the droplets. After 15 minutes, the temperature is reduced to 37ºC for 30 min, then reduced to 35ºC for 4 hrs with pressure and agitation unchanged, to fractionate the liquid into smaller droplets. Finally, the heating is stopped: the temperature decreases until the sol-gel transition is crossed, during which the agitation is maintained. Approximately 2 hrs later, the droplet suspension is quickly cooled to 15ºC with the serpentine, and left to set for 40 minutes with decreased agitation; the agitation is then stopped, and the pressure slowly reduced over 10 min to 11 bar, where it is maintained for ~ 15 hours with the temperature set to the selected measurement run temperature for the liquid.

Thereafter, the chamber pressure is slowly reduced to atmospheric, and the detector removed, weighed, and placed into either "cool" storage or utilization: high temperature implies an increased nucleation sensitivity, "cold" (< 0ºC) storage results in the formation of clathrate hydrates, which provoke spontaneous nucleation locally on the droplet surfaces in warming to room temperature, effectively destroying the device. Examples of the various completed fabrications are shown in Fig. 3.

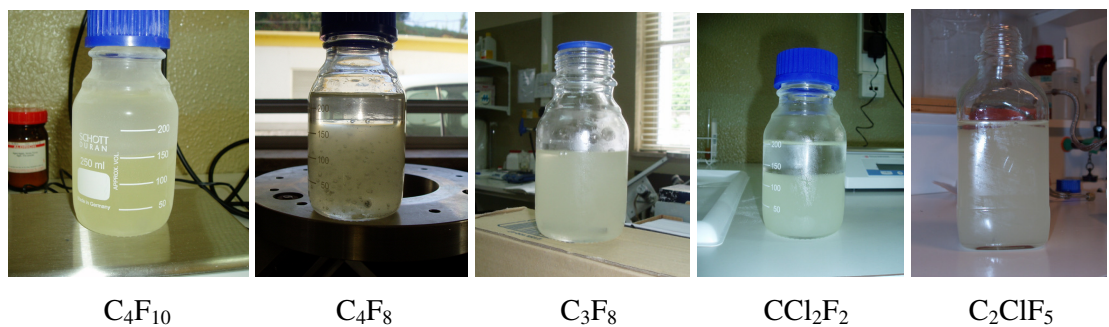

| $C_4F_{10}$ | $C_4F_8$ | $C_3F_8$ | $CCl_2F_2$ | $C_2ClF_5$ |

Fig. 3: examples of the various detector fabrications.

The agitation process fractionates the liquid droplets, resulting in a homogeneously-dispersed droplet size distribution: longer fractionating times generally give smaller diameter distributions; shorter times, larger distributions. The protocol is specific to the liquid, both in terms of time and speed. This is illustrated in Fig. 4, which presents fits of measured frequency distributions of droplet sizes in 5 μm intervals, obtained by optical microscopy from batch samples, for each of the SDDs with variations in their fractionating time and speed during their protocol development.

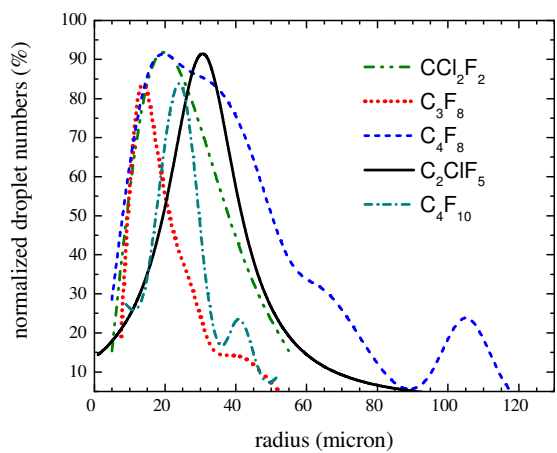

Fig. 4: various size distributions of fabricated detectors, resulting from variations in the fractionating time and speed, relative to a "standard" $C_2ClF_5$ fabrication.

## 2.3 Irradiation Tests

The laboratory "standard test" detector, a small version of the SIMPLE dark matter SDD fabricated with a scaled-down "standard" recipe protocol described above, contained ~ 2.7 g of $C_2ClF_5$ suspended in a gel matrix within a 150 ml laboratory bottle (Schott Duran GL45).

Similar SDDs were fabricated using the above "density-matched" protocol with $CCl_2F_2$ (2.5g), $C_3F_8$ (3.1g), and the "viscosity-matched" protocol with $C_4F_{10}$ (2.6g) and $C_4F_8$ (2.8g). None of the device gels were resin-purified in order to profit from the α decay of the intrinsic U/Th impurities. The fractionating time of each was adjusted to provide approximately identical, normally-distributed droplet sizes of $<r> = 30$ µm.

Once formed, each SDD was instrumented with the same capping used in the search experiments, a hermetic construction containing feedthroughs for a pressure line and a high quality electret microphone cartridge (Panasonic MCE-200) with a frequency range of 0.020–16 kHz (3 dB), SNR of 58 dB and a sensitivity of 7.9 mV/Pa at 1 kHz. The microphone, sheathed in a protective latex covering, was positioned inside the detector bottle within a 6 cm thick glycerin layer above the droplet emulsion, as shown in an empty device containment of Fig. 5: the microphone is seen below the cap, with the electronics cable interface vertical; the horizontal couple permits over-pressuring of the device up to 4 bar (the limit of the detector glass), and is coupled to a pressure transducer (PTI-S-AG4-15-AQ) for readout.

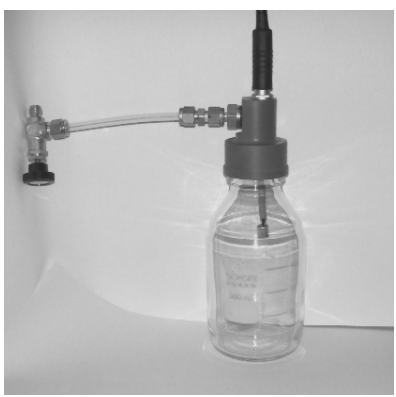

Fig. 5: empty detector, showing the microphone interface (vertical) and pressure couple (horizontal).

The microphone signal is remotely processed by a low noise, high-flexibility, digitally-controlled microphone preamplifier (Texas Instruments PGA2500), which is coupled to the archiving PC via an I/O board (National Instruments PCI-6251).

Once fabricated, each detector was placed in the same temperature-controlled water bath situated inside an acoustic foam cage designed for environmental noise reduction, despite the capability of the microphone-based instrumentation to distinguish between the various noise events [23]. Measurements were performed in steps of 5ºC over the temperature range of 5 - 35ºC. The temperature was measured with a type K thermocouple (RS Amidata 219-4450): each change was stabilized over ~ 20 minutes. Data was acquired in Matlab files of ~ 10 MB each at a constant rate of 32 kSps for periods of 5 minutes each. Nucleation events were generally stimulated by low level α radiation from the gel/glass U/Th impurities in order to provide time-separated events.

Figure 6 shows a typical, particle-induced bubble nucleation signal event, generally described as a damped sinusoid with a typical duration of several milliseconds, and its frequency spectrum in a standard $C_2ClF_5$ SDD. The Fast Fourier Transform (FFT) is characterized by a primary peak at ~ 640 Hz, with some lower power harmonics at ~2 and ~4 kHz. Non-particle induced signals have been well characterized in terms of their time constants (τ), amplitudes ($\mathcal{A}$) and frequencies ($\mathcal{F}$), and can be further discriminated from particle-induced events on the basis of their respective power density spectra which differ significantly from that of Fig. 6(b) [24].

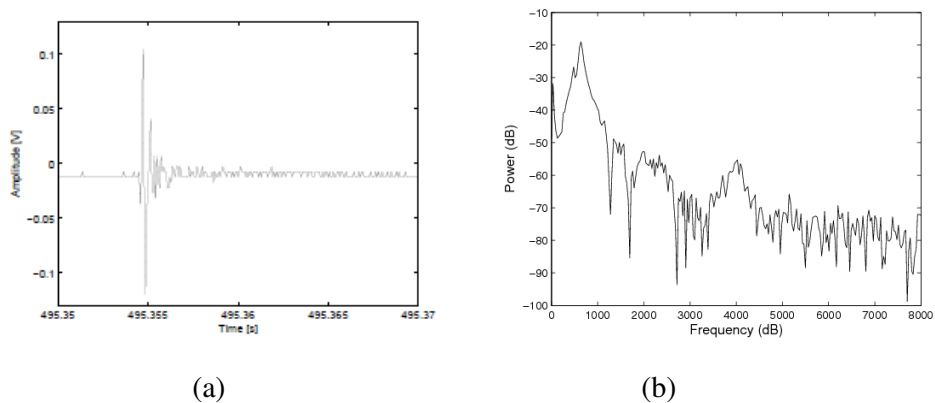

(a)          (b)

Fig. 6: typical instrumentation pulse shape (a) and FFT (b) of
a true particle-induced bubble nucleation event.

The results were subjected to a full, standard signal analyses [23]. The resulting acoustic background events were identified as normally-occurring gel fractures, trapped gas in the gel, and environmental noise intrinsic to SDD operation.

The noise levels were ~ 2 mV among all devices at all temperatures, except near 35°C where the level was ~ 4 mV since the detector gel was at a point of meltdown. A survey of the results at 1 bar is shown in Figs. 7(a)-(d); the 2 bar results will be discussed later. The error bars represent the standard deviation of the averages over the respective parameter measurement at each temperature: where not seen, they are smaller than the indicated data point.

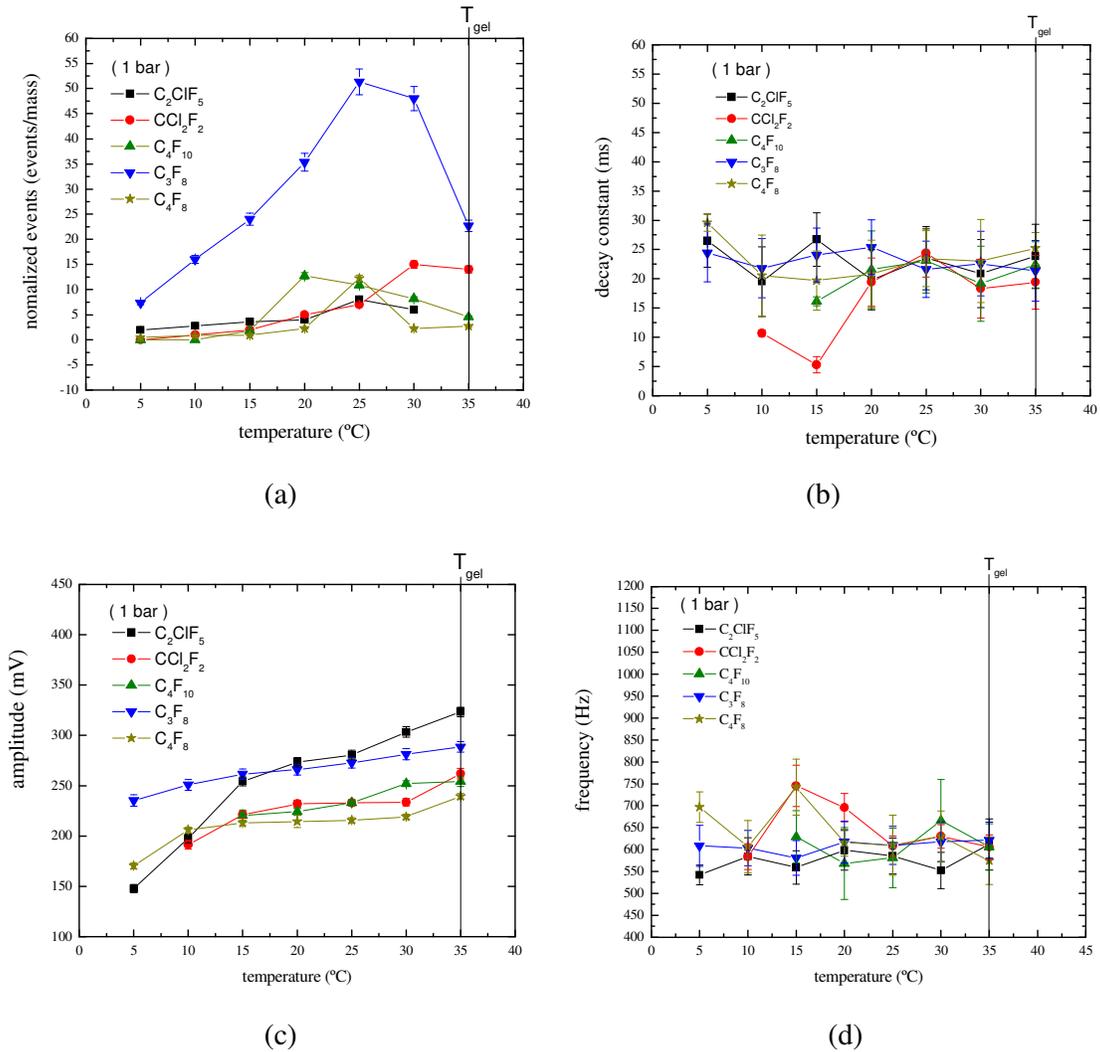

Fig. 7: nucleation response for different refrigerants at 1 bar: (a) event rates normalized to detector mass, and signal (b) $\tau$, (c) $\mathcal{A}$, (d) $\mathcal{F}$. The vertical line in each indicates the gel melt temperature ($T_{gel}$).

With the notable exception of the $C_3F_8$ event rates, the response of all liquids appears similar; with increasing temperature, the superheated liquids become more sensitive to incident radiation as a result of a reduced metastability barrier. Since the gel also becomes increasingly less stiff with temperature, an overall decreasing signal τ, increasing $\mathcal{A}$ and decreasing $\mathcal{F}$ might be expected. As seen in Figs. 7, all event rates tend to increase on approach to the gel melt temperature, as also the signal $\mathcal{A}$. In contrast, the signal τ's decrease, and $\mathcal{F}$'s fluctuate between 500–750 Hz. The results in all cases are consistent with the observed ranges observed with $C_2ClF_5$ for true bubble nucleations: τ within 5-40 ms, $\mathcal{F}$ within 0.45- 0.75 kHz [23]. The majority of signal $\mathcal{A}$ are > 125 mV: since neutrons in general produce nuclear recoil events with $\mathcal{A}$ < 100 mV [3], the results are consistent with the event triggering of the SDDs being principally from the α-emitting U/Th impurities of the detector gel and containment, as intended. Nonetheless, some events were recorded with $\mathcal{A}$ < 100 mV: 1 event with $CC_2F_2$ and 3 events with $C_4F_{10}$, to which we will return later.

The $C_3F_8$ device, in contrast to the other liquid SDDs, was a 2.1 wt% device, hence more susceptible to sympathetic nucleations occurring within the resolving time of the instrumentation. Also unlike the other devices, its gel above 30ºC was in a state of decomposition: the glycerine layer surrounding the microphone was filled with foam, and identification of a particle-induced signal increasingly difficult.

### 3. Superheated Liquids and Irradiation Response

In order to more fully appreciate the above results, we discuss several aspects of both the superheated liquids and their response to irradiations.

### 3.1 Superheated liquids

The physics of the SDD operation, the same as with bubble chambers and described in detail in Ref. [24,25] and references therein, is based on the "thermal spike" model of Seitz [26] which can be divided into several stages [27,28]. Initially, energy is deposited locally in a small volume of the liquid, producing a localized, high temperature region (the "thermal spike"), the sudden expansion of which produces a shock wave in the surrounding liquid. In this stage, the temperature and pressure of the liquid within the shock enclosure exceed the critical temperature and pressures, $T_c$ and $p_c$ respectively: there is no distinction between liquid and vapor, and no bubble. As the energy is transmitted from the thermalized region to the surrounding medium through shock propagation and heat conduction, the temperature and pressure of the fluid within the shock enclosure decrease, the expansion process slows and the shock wave decays. As the temperature and pressure reach $T_c$ and $p_c$, a vapor-liquid interface is formed which generates a protobubble. If the deposited energy was sufficiently high, the vapor within the protobubble grows to a critical radius $r_c$; if the energy was insufficient, cavity growth is impeded by interfacial and viscous forces and conduction heat loss, and the protobubble collapses.

To achieve $r_c$, the deposited energy (E) must satisfy two thermodynamic criteria:

$$E \geq E_c = 4\pi r_c^2 (\sigma - T\frac{\partial \sigma}{\partial T}) + \frac{4}{3}\pi r_c^3 \rho_v h_{lv} + \frac{4}{3}\pi r_c^3 \Delta p \qquad (1)$$

$$\frac{dE}{dx} \geq \frac{E_c}{\Lambda r_c}, \qquad (2)$$

where $r_c = 2\sigma/\Delta p$, $\sigma(T)$ is the droplet surface tension, $\Delta p = p_V - p$ is the liquid superheat, $p_V(T)$ is the vapor pressure of the liquid, p and T are the SDD operating pressure and temperature, $h_{lv}(T)$ is the liquid-vapor heat of vaporization, $\Lambda r_c$ is the effective ionic energy deposition length, and $\frac{E_c}{\Lambda r_c}$ is the critical LET. The first term represents the work required to create the protobubble interface; the second, the energy required to evaporate the liquid during protobubble growth to $r_c$. The third term describes the reversible work during protobubble expansion to $r_c$ against the liquid pressure. Generally, the second term is the largest, with the first ~ half. Not included in Eq. (1) are various irreversible processes which are generally small compared to the first three terms.

From Eq. (1), the $E_c$ for bubble nucleation is strongly dependent on the $h_{lv}$ of the liquid, the variation of which is shown in Fig. 8 for the various liquids investigated, as obtained from $h_{lv}(T) = \chi(1-T/T_c)^n$ with $\chi$ and n for each liquid shown in Table I, and all temperatures in K. As seen, $h_{lv}$ decreases with temperature increase.

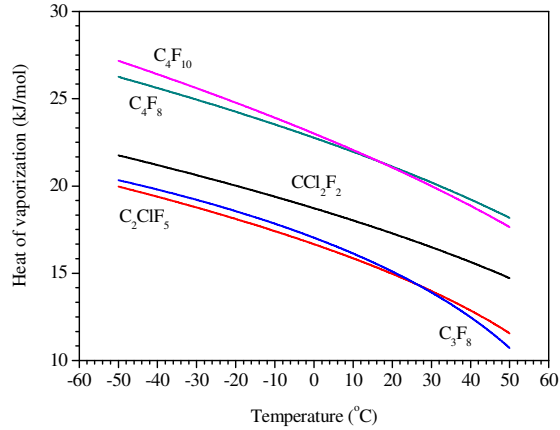

Fig. 8: variation of $h_{lv}$ with temperature for the various study liquids.

Table I: $\chi$, n for the study liquids, from Ref. [29].

| Liquid | $C_2ClF_5$ | $CCl_2F_2$ | $C_3F_8$ | $C_4F_8$ | $C_4F_{10}$ |
|---|---|---|---|---|---|
| $\chi$ (kJ/mole) | 28.99 | 30.93 | 30.67 | 36.82 | 41.21 |
| n | 0.373 | 0.406 | 0.383 | 0.396 | 0.455 |

The liquid response is also seen to depend on the nucleation parameter "$\Lambda$" of the liquid in Eq. (2), in effect defining the energy density required for bubble nucleation. Its variation with temperature is shown in Fig. 9, using $\Lambda = 4.3(\rho_V/\rho_l)^{1/3}$ which has been shown in agreement with experiment for $C_2ClF_5$ [22] and $CCl_2F_2$ [30]; although the $(\rho_V/\rho_l)^{1/3}$ is theoretically justified, its pre-factor is not in general and measurement is required.

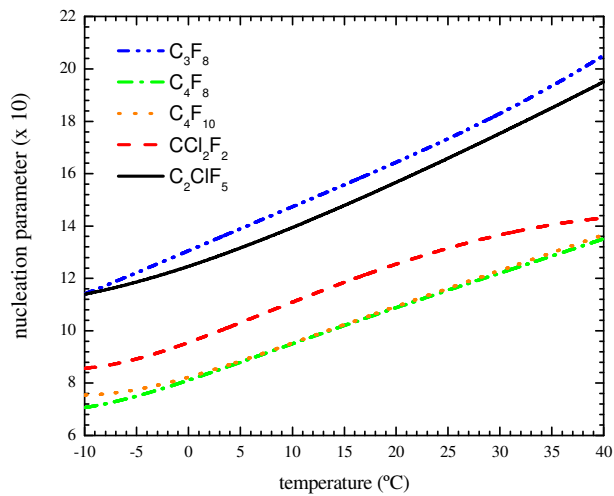

Fig. 9: variation of the nucleation parameter with temperature for various liquids, calculated with $\Lambda = 4.3(\rho_V/\rho_l)^{1/3}$.

The critical LET in each case, of order 100 keV/μm, is sufficiently high that bubble nucleations can be triggered only by high LET irradiations – either ion recoils generated by neutron scatterings or by α's.

The stopping power of the ions within the liquid is shown in Fig. 10(a) for the constituent nuclei of $C_2ClF_5$ and He from the U/Th contaminations of the SDD materials (which range in energy between 4.2 – 8.8 MeV) in $C_2ClF_5$ at 1.3 g/cm$^3$. Figure 10(b) displays the α threshold energy ($E_{thr}^{\alpha}$) of 5.5 MeV α's in $C_2ClF_5$ at 1 and 2 bar, calculated with Eqs. (1) and (2) using thermodynamic parameters taken from Refs. [19,31], α stopping powers calculated with SRIM 2008 [32], and the experimentally determined $\Lambda = 1.40$ for $C_2ClF_5$ at 2 bar and 9ºC [22,3]. The "nose" of the α curves in Fig. 10(b) reflects the He Bragg peak in $C_2ClF_5$ seen in Fig. 10(a), with the SDD sensitivity at a given temperature lying between the lower and upper contours. From Fig. 10(b), at 9ºC the α "window" thresholds are clearly reduced at higher pressure; at 2.2 bar, the "nose" lies to the right of the indicated 9ºC line, and the α sensitivity vanishes.

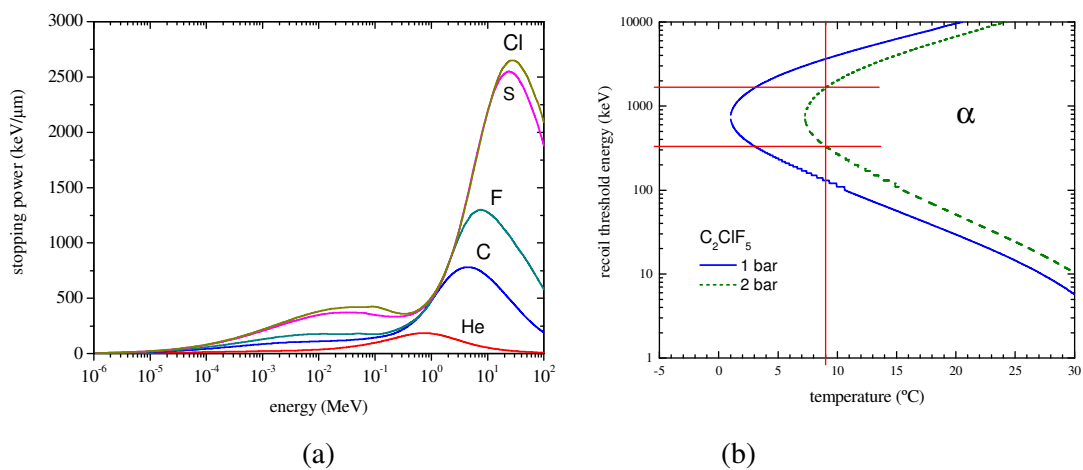

Fig. 10: (a) stopping power of α's and recoil ions in $C_2ClF_5$ (ρ=1.3 g/cm$^3$) as a function of energy; (b) variation of $E_{thr}^{\alpha}$ for $C_2ClF_5$ with temperature, at 1 (solid) and 2 (dashed) bar.

Figure 11 shows the elastic $E_{thr}^{nr}$ contours for each of the $C_2ClF_5$ constituents with temperature variation, calculated as for Fig. 10(b) using SRIM and the experimental $\Lambda = 1.40$. The curves reflect the respective constituent recoil ion stopping powers of Fig. 10(a), which for $E_{recoil} < 100$ keV are well-below the respective Bragg peaks.

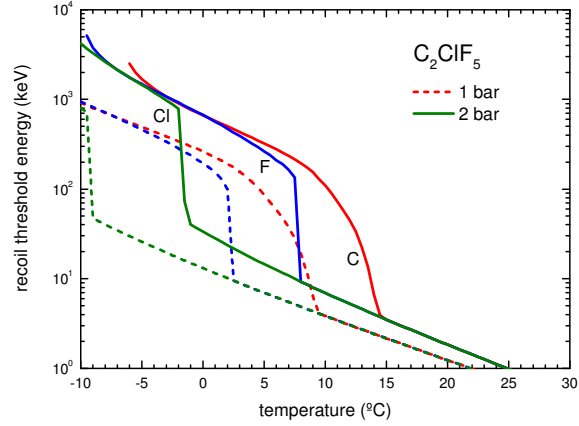

Fig. 11: variation of liquid ion threshold recoil energy curves for $C_2ClF_5$ with temperature, at 1 (dashed) and 2 (solid) bar.

The maximum ion recoil energy in a neutron elastic scattering on a nucleus of atomic mass A is given by $E_{recoil}^A = f_A E_n$, where $f_A = \frac{4A^2}{(1+A)^2}$ and $E_n$ is the incident neutron energy: for the liquids of this study, $f_F = 0.19$, $f_C = 0.27$ so that a detector with a minimum nuclear recoil threshold energy $E_{thr}^{nr} = 8$ keV implies a minimum response sensitivity to $E_n = 42$ and 30 keV for F and C recoils, respectively [45]. For the liquids with Cl, $f_{Cl} = 0.10$ and $E_{thr}^{nr} = 8$ keV implies $E_n = 80$ keV; there are also two inelastic reactions with $^{35}Cl$ which may induce events through their recoiling ions: $^{35}Cl(n,p)^{35}S$ and $^{35}Cl(n,\alpha)^{32}P$. In the first case the S ion has a maximum energy of 17 keV and can produce a nucleation for neutron energies $\leq 91$ keV, whereas the P ion emerges with a minimum energy of 80 keV that can always provoke an event. However, as these reactions have cross sections smaller than those of elastic scattering on Cl by ~ 1-7 orders of magnitude, their contribution to the detector signal is generally small (with exceptions in thermal neutron beams at reactors). For $C_2ClF_5$ above 15ºC, there is also the problem of events originating from high-dE/dx Auger electron cascades following interactions of environmental gamma rays with Cl atoms in the refrigerant.

The metastable barrier decreases with increasing temperature, which is by virtue of $f_A$ sequentially overcome by the recoiling constituent ions until a common threshold is reached at ~ 10ºC (15ºC) at 1 (2) bar. At 9ºC and 2 bar, $E_{thr}^{Cl,F} = 8$ keV while $E_{thr}^{C} \sim 80$ keV. For fixed temperature operation, SDD pressure increase raises the $E_{thr}^{nr}$ curve and shifts it to higher temperatures.

As evident, the response sensitivity of each liquid is not the same at each temperature. This is a result of the variation in degree of superheating of the liquids, which varies significantly with T and p as seen in Table II. A "universal" characterization of the response is obtained by replacing the temperature with the reduced superheat factor, $S = (T - T_b)/(T_c^* - T_b)$ with $T_b$ the boiling temperature of the liquid at a given pressure and the critical temperature $T_c^* = 0.9T_c$, with all temperatures in K, since the fluid phase of organic liquids ceases to exist at a temperature about 90% of the tabulated critical temperature $T_c$ [33]. Equations (1) and (2), when satisfied simultaneously, provide the threshold energy ($E_{thr}$) for bubble nucleation, which when displayed as a function of S fall on a "universal" curve for the nucleation onset of superheated liquid devices [25]. The range of S for each liquid is also shown in Table II. Numerous studies have shown the insensitivity of various liquid devices to γ's, cosmics and minimum ionizing radiations for S < 0.72 [25,34].

Table II: critical ($T_c$) and boiling temperature ($T_b$) at 1 and 2 bar for the different liquids (data from NIST [19]).

| Temp./Refrig. | $C_2ClF_5$ | $CCl_2F_2$ | $C_4F_{10}$ | $C_3F_8$ | $C_4F_8$ |
|---|---|---|---|---|---|
| $T_c$ (ºC) | 79.95 | 111.97 | 113.18 | 71.95 | 115.23 |
| $T_b$ (ºC) – 1 bar | -38.94 | -29.75 | -2.09 | -36.83 | -5.98 |
| S (5-35ºC) – 1 bar | 0.52-0.88 | 0.34-0.63 | 0.09-0.48 | 0.56-0.97 | 0.13-0.50 |
| $T_b$ (ºC) – 2 bar | -22.15 | -12.13 | 16.56 | -20.20 | 12.13 |
| S (5-35ºC) – 2 bar | 0.41-0.86 | 0.20-0.55 | 0-0.32 | 0.44-0.96 | 0-0.36 |

## 3.2   The Irradiation Test Results Revisited

Given the above considerations, we now display the full experiment results with respect to S, beginning with the device responses in Fig. 12. Since the gel melting temperature ($T_{gel}$) is absolute hence appears at a different S for each liquid, these (denoted $S_{gel}$) are indicated for each device throughout.

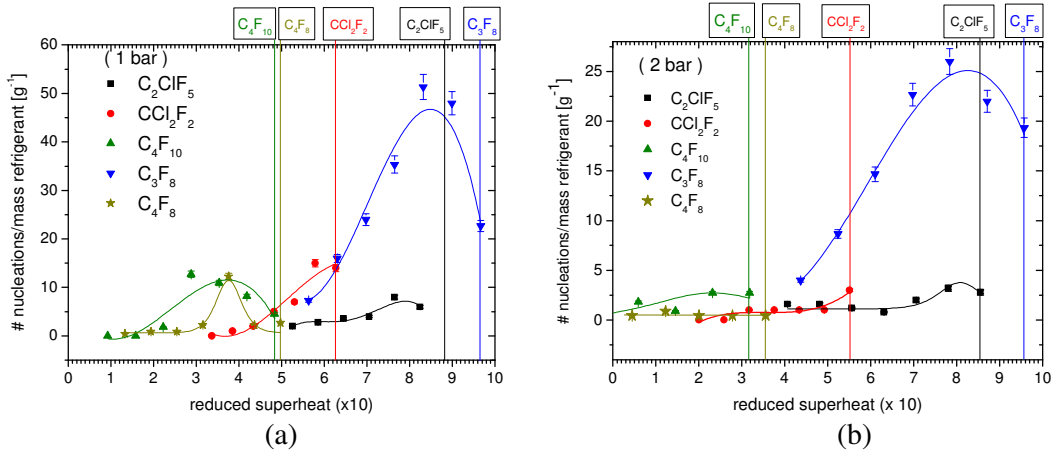

Fig. 12: nucleation response for different refrigerants at (a) 1 and (b) 2 bar. The identified lines indicate the S for each liquid associated with $T_{gel}$.

A higher reduced superheating implies a lower metastable energy barrier: the general response should be an asymmetric sigmoid, with the onset of minimum ionizing events occurring at S ~ 0.7. For S < 0.7, all event numbers should generally remain flat or increase with temperature depending on the degree of superheating, as observed herein; for S > 0.7, the liquids are increasingly sensitive to lower LET radiations which provide an additional contribution to the event rates.

In the case of $C_3F_8$, with an event response a factor ~ 10 larger than the other devices, the liquid above S = 0.8 is near its foam limit (S = 1) at which vapor phase transitions occur via thermal fluctuations, providing an explanation for the observed gel conditions (see Sec. 2.3). Moreover, its $E_{thr}^{\alpha}$ is near 10 keV, an order of magnitude below that of the other liquids hence more susceptible to α's otherwise reduced in energy by the gel to below thresholds of the other liquids. Apart from this, the geometric efficiency for α-induced nucleations increases for small droplet sizes as $\varepsilon = 0.75 f R_\alpha/r$, where f is the active mass fraction, $R_\alpha$ is the alpha particle range in the liquid, and r is the droplet radius [34]. Fig. 4, the $C_3F_8$ device exhibits the smallest size distribution of all, with <r> = 15 ± 9 μm; using the f's of the prototype fabrications, $\varepsilon_{C3F8}/\varepsilon_{C2ClF5}$ ~ 3.5 consistent with Fig. 12.

In both regimes of S, the response is moderated by the effects of the gel becoming increasingly less stiff as its melting regime is entered [20]. The observed decrease of the $C_3F_8$ event response below $S_{gel}$ is also in evidence for the other liquids, all of which are in states of S < 0.7, and well below their respective foam limits, suggesting the gel relaxation to be principally responsible for the decrease.

The remainder of the results relate to the microphone-recorded signal characteristics which result from a bubble nucleation event, with the τ, $\mathcal{F}$ and $\mathcal{A}$ of the particle-induced signal events for each liquid as a function of S at each pressure shown in Figs. 13-15, respectively.

Figures 13 display the signal decay constants: as anticipated, all are generally contained within 10-30 ms, with most showing an increase with temperature as a result of decreasing gel stiffness. With the exception of $C_3F_8$, all τ in Fig. 13(a) initially manifest considerable dispersion, condensing to 20-25 ms by 20ºC; $C_3F_8$ shows a slight decrease with approach to $S_{gel}$. In Fig. 13(b), the τ of $C_4F_8$ increases on approach to $S_{gel}$, then drops to 10 ms thereafter; for both $C_4F_{10}$ and $CCl_2F_2$, the τ fluctuates between 10-30 ms. In contrast, the $C_3F_8$ and $C_2ClF_5$ signal τ remain generally unchanged with temperature increase. Note that the τ of the 2 bar results are generally slightly increased relative to the 1 bar results, again as might be expected from a stiffer gel [20]. Also note, from Ref. [23], that τ's for non-particle induced events are generally > 36 ms.

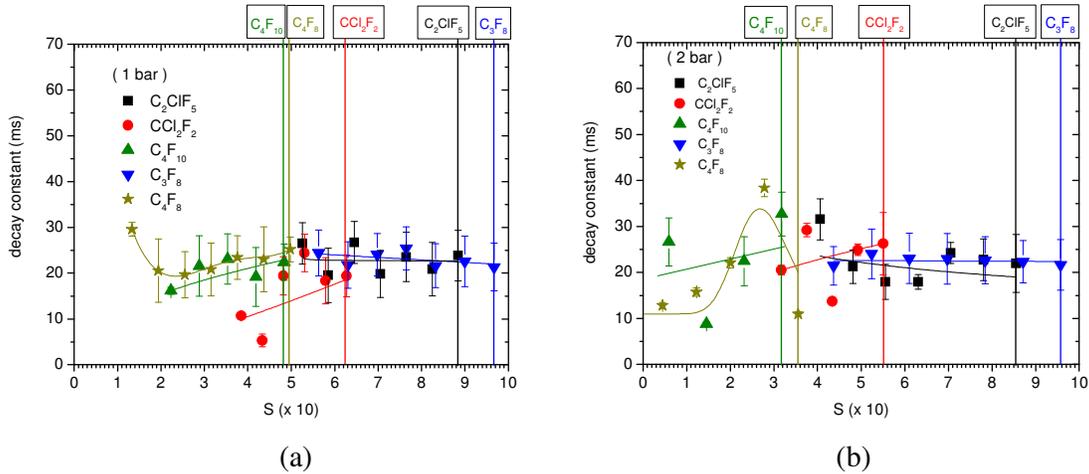

Fig. *13: signal* τ for different refrigerants at (a) 1 and (b) 2 bar.

Since the frequency of an event is also defined by the elasticity of the medium, the signal $\mathcal{F}$ should tend to decrease with increasing temperature. This is not immediately discernible in Figs. 14. At 1 bar, $\mathcal{F}$ generally fluctuates until ~30ºC (possibly the result of low statistics), with $C_2ClF_5$ showing an increase in approach to $S_{gel}$. At 2 bar, the $\mathcal{F}$ are slightly lower and more dispersed than those at 1 bar, with $C_3F_8$ also showing an increase towards $S_{gel}$. Note that the recorded $\mathcal{F}$ differ significantly from those reported by PICASSO and COUPP, and that $\mathcal{F}$ of acoustic background events are generally < 100 Hz [23].

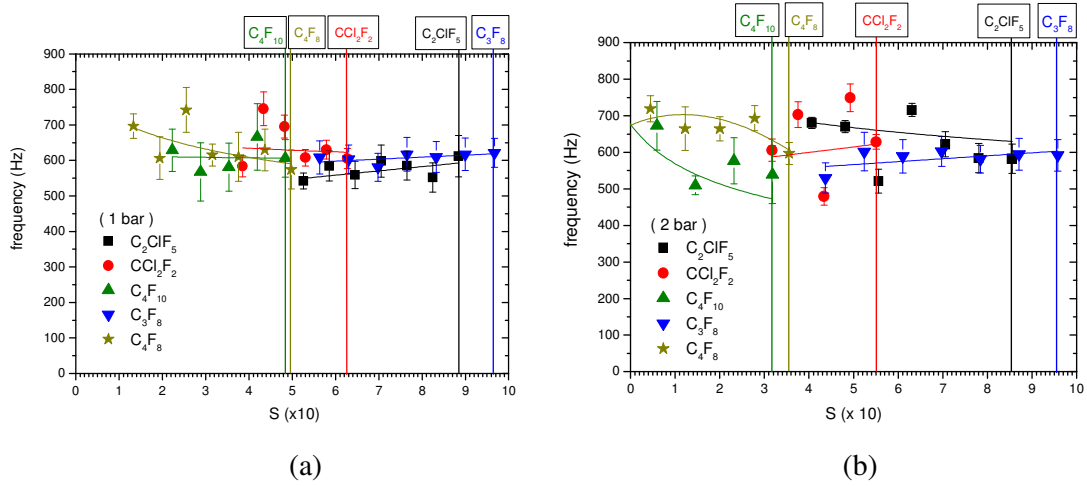

Fig. 14: signal $\mathcal{F}$ variations for different refrigerants at (a) 1 and (b) 2 bar.

The complete phase transition of a droplet results in a gas bubble harmonically oscillating about its equilibrium radius $r_b$. The resonant frequency is given by Minnaert [39]:

$$f_r = \frac{1}{2\pi r_b}\sqrt{\frac{3\eta p_0}{\rho_l}} \quad , \quad (3)$$

where $\eta$ is the polytropic coefficient of the gas, $p_0$ is the ambient equilibrium pressure (effects of bubble movement caused by buoyancy forces, and spatial variation of the pressure during the growth process are neglected [40]). For typical parameters at 9ºC and 2 bar, $r_b$ = 5 mm and $\eta \sim 1.1$ [41], $f_r = 700$ s$^{-1}$ consistent with the event records.

As seen in Figs. 15, all $\mathcal{A}$ generally increase with approach to $S_{gel}$, as expected with increased superheating, and are generally lower at 2 bar than those at 1 bar, as also expected with a stiffer gel.

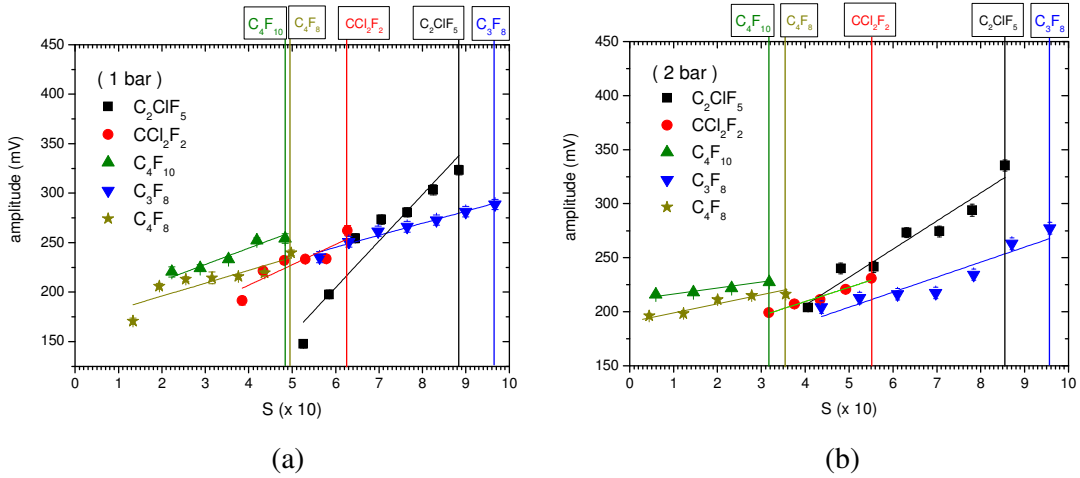

Fig. 15: $\mathcal{A}$ variations for different refrigerants at (a) 1 and (b) 2 bar.

The emitted acoustic power (J) in a bubble nucleation is proportional to the acceleration of the bubble volume expansion [42],

$$J \sim \frac{\rho_1}{4\pi c} \ddot{V}^2 \ , \qquad (4)$$

where c is the speed of sound in the medium, V is the droplet volume and the dots denote differentiation with respect to time. The pressure P produced in a liquid bath without gel at a distance d from the source at time t is then $\frac{1}{d}\sqrt{\frac{c\rho_1}{4\pi}J}$ which with $V = \frac{4\pi}{3}r^3$ reduces to

$$P(d,t) = \frac{\rho_1}{4\pi d}\ddot{V} = \frac{\rho}{d}\left[2r\dot{r}^2 + r^2\ddot{r}\right] \ . \qquad (5)$$

An idea of the pressure change is obtained from the solution to the Rayleigh-Plesset equation in the asymptotic limit [43]:

$$r(t) = t\left[\frac{2}{3\rho_1}\Delta p\right]^{1/2} \equiv t \cdot v_0(T) \ , \qquad (6)$$

such that $\dot{r} = v_0$, $\ddot{r} = 0$ and Eq. (5) becomes $P = \frac{2t\rho_1}{d}v_0^3$. Figure 16 displays the calculated temperature variation of $v_0$ for the liquids at 1 and 2 bar: note that all are continuously increasing, and similar to the experimental measurements in Fig. 15.

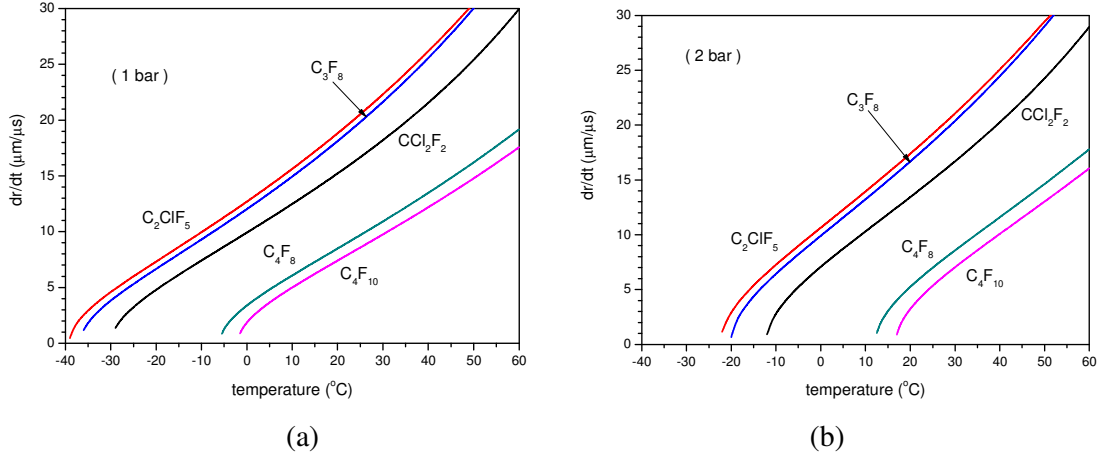

Fig. 16: variation of calculated $v_0$ with temperature, for (a) 1 bar, and (b) 2 bar.

At 9ºC and 2 bar, Fig. 16 gives $v_0$ ($C_2ClF_5$) ~ 13 µm/µs. Transducers respond to P changes with sensitivities of µV/µbar, with the sensitivity of the MCE-200 quoted at 7.9 mV/Pa at 1 kHz (±2dB) [44]: for $C_2ClF_5$, P ~ 6.2 x $10^2$ µbar over the first 1 µs at a distance of 10 cm, yielding signal $\mathcal{A}$ of ~ 1000 mV in reasonable agreement with those recorded experimentally for all liquids at all pressures.

### 3.3 Dark Matter Sensitivities

All direct dark matter search efforts are based on the detection of nuclear recoil events generated in WIMP-nucleus elastic scatterings. Neutrons also produce single recoil events via elastic scattering, generating a signal which is indistinguishable from that of WIMPs, and the response characterization of a detector to such recoils is generically obtained from neutron calibration measurements, either via weak neutron sources such as Am/Be or $^{252}$Cf, or the use of accelerator or reactor facilities which provide monochromatic neutron beams.

The calculated variation in the minimum $E_{thr}^{nr}$ for both pressures for the various liquids of this study, using $\Lambda = 4.3(\rho_V/\rho_l)^{1/3}$, is shown in Fig. 17. Note the group separation which reflects the respective liquid densities: the higher density liquids must be operated at higher temperatures or lower pressures to achieve the same threshold as the lower density liquids; for example, $C_4F_{10}$ operated at 1 bar and ~ 42ºC provides the same $E_{thr}^{nr}$ as $C_2ClF_5$ at 9ºC or $CCl_2F_2$ at 29ºC when operated at 2 bar.

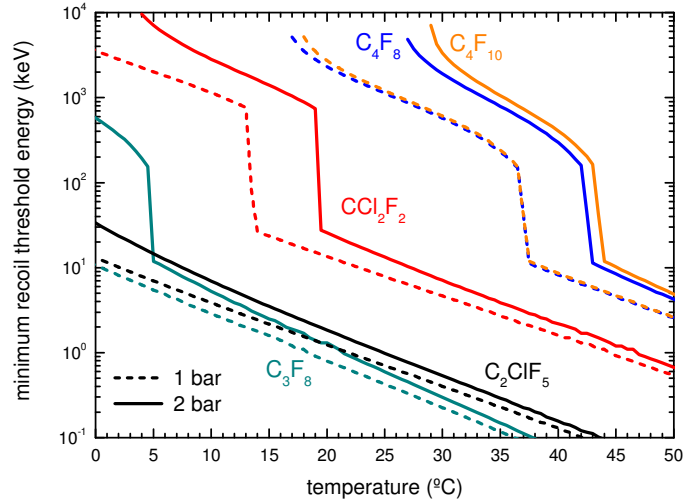

Fig. 17: variation of $E_{thr}^{nr}$ for $CCl_2F_2$, $C_2ClF_5$, $C_4F_{10}$, $C_3F_8$ and $C_4F_8$ with temperature, at 1 (dashed) and 2 (solid) bar.

While $C_4F_8$ and $C_4F_{10}$ only provide recoil $E_{thr}^{nr} < 8$ keV at temperatures $> T_{gel}$ at either pressure, $C_3F_8$ permits a recoil $E_{thr}^{nr} \sim 2.4$ keV at 15ºC and 2 bar, and $C_2ClF_5$ a recoil $E_{thr}^{nr} \sim 5$ keV at 12ºC and 2 bar; operation at 12ºC and 1 bar provides $E_{thr}^{nr} = 3$ keV. Lower overpressuring of the SDDs generally provides lower $E_{thr}^{nr}$, but operation at 2 bar is preferred as a radon suppression measure.

Apart from $E_{thr}^{nr}(T,P)$, the quality of any search effort depends on the detector's active mass, exposure and target sensitivity. The liquid solubility determines the amount of active target mass in the detector, as well as the fracture probability of the gel. Although fracturing occurs with or without bubble nucleation, since the liquid occupies any microscopic $N_2$ gas pockets formed during the fractionating stage of the suspension fabrication, it is aggravated by nucleations arising from the ambient background radiations of the fabrication site. Table III displays the solubilities of the various liquids. As seen, the solubility of $C_4F_8$ is ~ half of $C_2ClF_5$, with $C_3F_8$ a factor 4 lower; $C_4F_{10}$ is the least soluble by a factor of ~ 10 relative to $C_2ClF_5$. Of all, $CCl_2F_2$ is the most soluble, hence may easily suffer from a reduced, time-dependent active liquid concentration and lower triggering probability. These numbers however vary significantly between compilations, and must be measured for each liquid and suspension material prior use.

Table III: liquid solubilities in water (in g/liter/bar) at 25ºC, together with the active freon mass of each prototype detector in the reported measurements.

|  | $C_2ClF_5$ | $CCl_2F_2$ | $C_3F_8$ | $C_4F_8$ | $C_4F_{10}$ |
|---|---|---|---|---|---|
| Solubility | 0.058[35] | 0.28[36] | 0.015[37] | 0.025[37] | 0.005[38] |
| Active mass (g) | 2.7 | 2.5 | 3.1 | 2.8 | 2.6 |

Prior 2005, SIMPLE SDDs with $C_2ClF_5$ were usable for ~ 40 day as a result of signal avalanches resulting from fracture events [22], which the early fabrication chemistry did not address and the instrumentation was unable to discriminate. The lifetime has effectively increased to ~ 100 day, largely via instrumentation improvements which permit identification of fracture events, but also with improvements in the gel/detector fabrication to include the use of PVP and agarose, prohibition of storage below 0ºC, and on-site detector fabrications in a quasi-clean room environment. Measurements conducted in 2006-2007 with a $C_2ClF_5$ SDD indicated an abrupt increase in the measured noise level only after 109 d of operation as a result of massive fracturing.

Apart from the liquid response, its dark matter search sensitivity depends on its constituent target A and spins. The WIMP-nucleus cross section $\sigma_A$ is to first order a sum of spin-independent (SI) and spin-dependent (SD) contributions, $\sigma_A = \sigma_{SI} + \sigma_{SD}$, with

$$\sigma_{SD} = \frac{32}{\pi} G_F^2 \mu_A^2 \left[ \left(a_p \langle S_p \rangle + a_n \langle S_n \rangle\right)^2 \frac{J+1}{J} \right] \quad , \quad (7)$$

$$\sigma_{SI} = \frac{4}{\pi} G_F^2 \mu_A^2 \left[g_p Z + g_n N\right]^2 \quad , \quad (8)$$

with $G_F$ the Fermi constant, $g_{p,n}$ ($a_{p,n}$) the SI (SD) WIMP couplings with the proton (neutron) respectively, $\mu_A$ the WIMP-nuclide reduced mass, and J the total nuclear spin. With isospin conservation, $g_p = g_n = 1$ and $\sigma_{SI} \sim A^2$ : in comparison with the Xe-based experiments for example, the heavier target result is enhanced by a factor of $(131/19)^2 = 48$. Since fluorine possesses the largest $<S_p>$ of all nuclides in common use ($<S_p>$ = 0.475 [46]), superheated liquids have generally provided the most sensitive target for WIMP-proton SD studies, with less impact in the SI sector relative to their heavier counterparts owing to the $A^2$ enhancement of the WIMP-nucleus cross section.

As an example, consider $C_3F_8$ relative to the recent SIMPLE $C_2ClF_5$ result [3]: the effect of the larger fluorine component of $C_3F_8$ in the SD sector is shown in Fig. 18(a), assuming identical measurement results. As seen in Fig. 18(b) however, in the SI sector, the $C_3F_8$ impact is severely weakened, despite a molecular mass of 198 vs. the 154 of $C_2ClF_5$. A fictional "$C_3ClF_8$" liquid yields a contour almost identical to $C_2ClF_5$, with the difference attributed to the Cl mass fraction of the molecule (0.17 for $C_3ClF_8$ vs. 0.24 for $C_2ClF_5$). Each exclusion calculation includes the C presence, suggesting its "spectator" presence in the measurement.

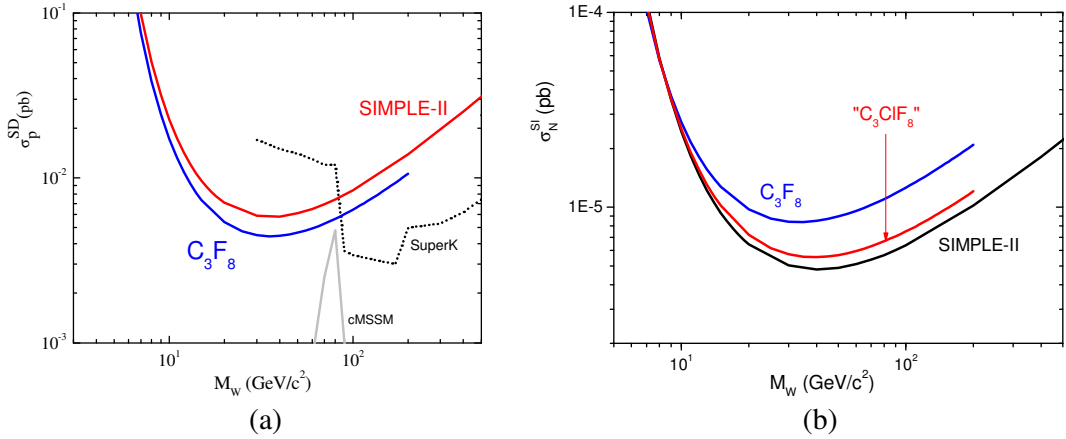

Fig. 18: comparison of $C_3F_8$ sensitivities with $C_2ClF_5$ in both the spin-dependent (a) and spin-independent (b) sectors, for identical exposure and sensitivity.

Light nuclei liquids may still contribute to the SI sector because of the low recoil threshold energies possible with the technique, since the low $M_w$ part of the exclusion contour tends to flatten with decreasing recoil $E_{thr}$ [47] as seen in Fig. 19 for $C_4F_{10}$ with a 121 kgd exposure and no observed candidate events. Over an order of magnitude improvement in experimental sensitivity at low WIMP mass derives from a reduction in $E_{thr}^{nr}$ from 16 to 6 keV. This is also observed in Ref. [3], where SIMPLE at $E_{thr}^{nr}$ = 8 keV all but eliminates the CoGeNT result [48] at low $M_w$, while COUPP at $E_{thr}^{nr}$ = 15 keV – although more sensitive at higher $M_w$ [2] -- is unable to contribute.

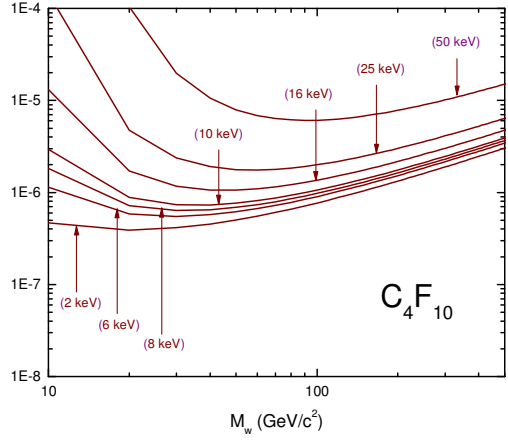

Fig. 19: variation of the $C_4F_{10}$ exclusion contour in the SI sector with decrease in the measurement recoil $E_{thr}^{nr}$ as indicated for a 121 kgd exposure with no candidate events.

## 4. Particle Discrimination

Fundamentally, the ability of any detector to contribute to a dark matter search depends on its capability to discriminate between nuclear recoil and background α-induced signals, as demonstrated by all three superheated liquid programs on the basis of their respective signal $\mathcal{A}$. In contrast to PICASSO and COUPP, in which the A obtain from integrations of the measured FFTs over a broad frequency range, the SIMPLE discrimination derives solely from the existence of a 30 mV gap between the recoil and α-induced event distributions of the primary harmonic of the FFTs, as seen in Fig. 20(a) [49]. This defines an empirical gap criterion of $\mathcal{A}_\alpha^{min} > \mathcal{A}_{nr}^{max}$.

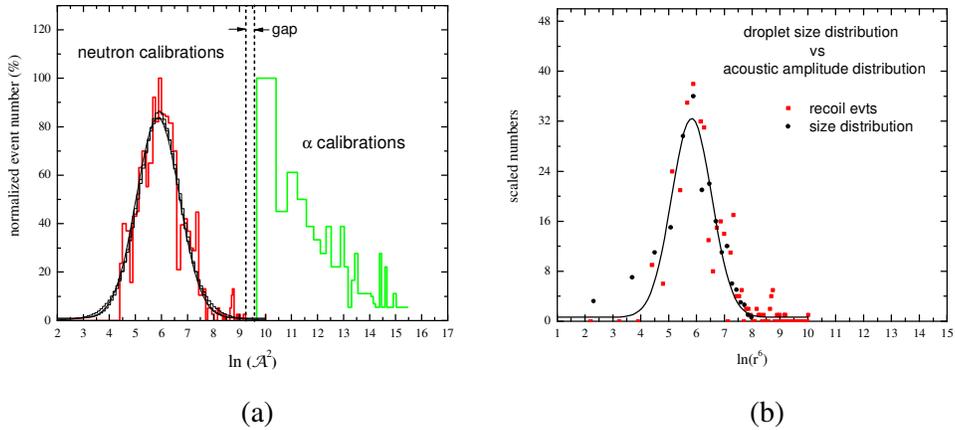

(a)          (b)

Fig. 20 : (a) initial neutron – α discrimination as reported in Ref. [49];
(b) overlap of droplet size distribution with nuclear recoil events.

Although there is to date no complete understanding of this discrimination in any of the programs, the consensus is that its principle origin lies in the difference between the energy loss of the α and recoil interactions within the liquid, and inherent protobubble formation. In general, a recoil event is the result of a single neutron elastic scattering interaction anywhere in a droplet, in which the LET of the recoil ion exceeds the critical LET for bubble nucleation only within a micron of the scattering origin in the liquid as shown in Fig. 21(a): only O(1) protobubbles can be formed. The recoil event distribution mirrors the droplet size distribution, as indicated in Fig. 20(b) with the solid contour representing the normalized $\ln(r^6)$ distribution of Fig. 4 and a shift to match the means. From Eq. (4) with $V = \frac{4\pi}{3}r^3$, and $t_0$ a characteristic single protobubble nucleation time, $\mathcal{A}_{nr}^{max} \sim r_{max}^3 t_0^{-2}$.

In contrast, the region of α LET > critical is generally distributed over several microns in the liquid, as seen in Fig. 21(b), so that an α event is capable of generating a number of protobubbles ($n_{pb}$); since each protobubble constitutes an evaporation center for the droplet, $t_\alpha = t_0/n_{pb}$ and $\mathcal{A}_\alpha^{min} \sim n_{pb}^2 t_0^{-2} r_{min}^3$ -- $n_{pb}^2$ constitutes an amplification factor for the α-generated amplitudes.

Thus the gap criterion reduces to $n_{pb}^2 r_{min}^3 > r_{max}^3$. Consider for example the energy loss of 5.5 MeV α's in $C_2ClF_5$ at 9ºC and 2 bar shown in Fig. 21(b): the critical LET (176 keV/μm) is only exceeded at penetration depths of 34-40 μm, with an estimated $n_{pb} \sim 12$ per micron. Ignoring for the moment the PVP presence in the gel, the α's originate from the droplet-gel interface [50], and droplets with r < 17 μm cannot form a protobubble (providing an effective lower cutoff ($r_{min}$) to the observed $\mathcal{A}_\alpha$ spectrum [51,3]: with $r_{max}$ = 60 μm from Fig. 4, the gap criterion is satisfied, and particle discrimination may be anticipated. For $E_\alpha$ = 8 MeV, the cutoff increases to $r_{min} \sim 33$ μm since the Bragg peak is translated to larger penetration depths, and the criterion is easily satisfied.

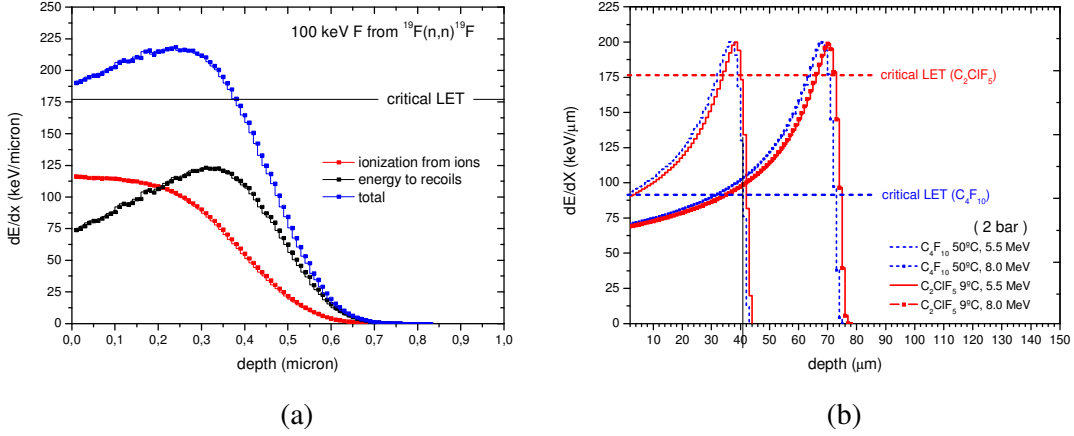

Fig. 21: (a) LET of recoil F ions in $C_2ClF_5$ as a function of penetration depth; (b) comparison of the energy loss depth profiles for 5.5 and 8.0 MeV α's in $C_2ClF_5$ and $C_4F_{10}$, at 2 bar and representative low $E_{thr}$ temperatures.

The situation differs in the case of $C_4F_{10}$, since the critical LET at 2 bar and 50ºC is only 103 keV/μm, which as seen in Fig. 21(b) is exceeded by 5.5 MeV α penetrations between 9-41 μm. This suggests $r_{min}$ ~ 5 μm: with $n_{pb}$ ~ 12 per micron, $\mathcal{A}_\alpha^{min} < \mathcal{A}_{nr}^{max}$, and $\mathcal{A}_\alpha$'s should be mixed with $\mathcal{A}_{nr}$, as in fact observed in the measurements herein which yielded 3 events with $\mathcal{A}_\alpha < 100$. While this is not the case for $E_\alpha$ = 8.0 MeV (where $r_{min}$ ~ 21 μm), in order to achieve full exclusion of the U/Th contaminant α-contributions in a dark matter search, the $C_4F_{10}$ droplet size distribution would likely need to be reduced to <r> ~ 5 μm.

The situation differs at 1 bar operation, where the critical LET for $C_2ClF_5$ and $C_4F_{10}$ are 123 keV/μm and 70 keV/μm, respectively. In this case, for $C_2ClF_5$, $r_{min}$ ~ 11 μm and the gap criterion is unsatisfied, as also for $C_4F_{10}$ with $r_{min}$ ~ 0.

We stress that the critical LET is dependent on Λ, which is not well-known in the case of $C_4F_{10}$ or most other liquids of this study, as well as the estimate of $n_{pb}$ which varies for each liquid, and that the above illustration neglects entirely non-interface α origins (although the particle LETs in gel are insignificantly different from the liquids). The PICASSO-determined Λ = 3.8 for $C_4F_{10}$ at 1 bar and 24ºC [52] is however higher than the $4.3(\rho_V/\rho_l)^{1/3}$ estimate of 1.13, and would lower the critical LET, worsening the situation. Although the PVP presence in the gel fabrication acts in part to suppress heavy ion migration to the droplet-gel interface, the efficiency is evidently < 100%. Further study is required to provide a complete description of the gap formation, and the particle discrimination capacity of each SDD must therefore be determined both experimentally and individually.

## 5. Heavier Nuclei Liquids

Given the above, one might immediately question whether SDDs using fluorine-based liquids with heavier A nuclei in detector fabrications are possible, towards maximizing a single experiment sensitivity in both SD and SI sectors. The question is not new, being in part responsible for the use of $CF_3I$ by COUPP. Because of its place in the periodic table, fluorine combines well with a variety of heavier halogens, offering a large number of possibilities which would provide the desired $A^2$ enhancement in the SI sector, to include I (IF, $IF_3$, $IF_5$, $IF_7$), Xe ($XeF_2$, $XeF_4$, $XeF_6$), Te ($TeF_5$), Ta ($TaF_5$), W ($WF_6$), Re ($ReF_6$) and a variety of fluorocarbons ($CF_3I$, $CBrF_3$, $CBrClF_2$,…) – in most cases, with the heavier nuclei constituents possessing sufficient $<S_{p,n}>$ [53-55] for significantly contributing in the SD sector as well; in the cases of Xe, Te, and W, the predominant contribution would be in $<S_n>$, simultaneously with the $<S_p>$ of fluorine.

The immediate considerations to be addressed are: (1) fabrication feasibility of a quality SDD, and (2) dark matter search sensitivity. An immediate caveat, following from the light liquids, is that the higher the density, the generally higher are the recoil thresholds and solubilities (e.g. those of $IF_5$ and $IF_7$, 0.8 g/liter and 0.5 g/liter respectively, are significantly higher than $C_2ClF_5$). A cursory overview of the possible candidates moreover indicates that none of the Xe compounds are liquids at temperatures usable with SIMPLE gels; $XeF_6$ is liquid in the window of 49-76ºC [56] and hydrolytic; $UF_6$ reacts with water, and $ClF_5$ is corrosive;

Generally, however, little is known regarding the liquid phase of such possibilities, in particular the thermophysical properties necessary to calculation of their respective $E_c$. Before embarking on an investigation of the properties, which would in most cases require dedicated measurements, it's useful to consider some screening of possible choices as regards their dark matter search suitability using the lessons obtained with the light nuclei liquids above.

## 5.1 Liquid Selection

As seen from Fig. 17 and the definition of S, a figure of merit for the recoil threshold energies can be defined by FM = T (S = 0.7), the temperature at which the liquid S = 0.7: the lower the FM, the lower the recoil threshold. We show in Table IV a small compendium of FMs for a number of heavy liquids possibilities, together with known thermophysical data and – following the discussion of Sec. 3.3 – the heavy nuclei mass fractions (see Sec. 4).

Table IV: FM's of various possible heavy target nuclei.

| Liquid | $\rho$ (g/cm3) | heavy mass fraction | $T_c$ (°C) | $T_b^{1\ bar}$ (°C) | FM (°C) |
|---|---|---|---|---|---|
| $ClF_5$ | 1.9 | 0.27 | $143^{57}$ | $-13^{57}$ | 69.4 |
| $BrF_5$ | 2.47 | 0.46 | $197^{58}$ | $41^{58}$ | 119.5 |
| $SF_6$ | 1.68 | 0.22 | $45^{59}$ | $-64^{59}$ | -8.43 |
| $MoF_6$ | 3.5 | 0.46 | $278^{59}$ | $34^{59}$ | 170. |
| $TeF_6$ | - | 0.52 | $107^{59}$ | $-38^{59}$ | 39. |
| $XeF_6$ | 3.56 | 0.53 | $229^{60}$ | $46^{56}$ | 142 |
| $WF_6$ | 3.43 | 0.62 | $178^{59}$ | $18^{59}$ | 101 |
| $ReF_6$ | 6 | 0.62 | $297^{60}$ | $34^{59}$ | 182 |
| $PtF_6$ | 3.83 | 0.63 | $93^{59}$ | $69^{59}$ | 59.9 |
| $UF_6$ | 5.1 | 0.68 | $230^{59}$ | $56^{59}$ | 145 |
| $CF_3I$ | $2.0^{58}$ | 0.65 | $123.3^{61}$ | $-21.83^{61}$ | 54.1 |
| $CBrF_3$ | $1.53^{56}$ | 0.54 | $66.93^{62}$ | $-57.79^{62}$ | 7.52 |
| $CBrClF_2$ | $1.8^{56}$ | 0.48 | $153^{63}$ | $-4.0^{63}$ | 78.3 |

Clearly, $SF_6$, $CBrF_3$, $TeF_6$, $CF_3I$, $ClF_5$ and $CBrClF_2$ (in descending order) provide the lowest threshold, whereas $UF_6$, $CF_3I$, $PtF_6$, $WF_6$ or $ReF_6$, and $CBrF_3$ provide the larger mass fractions, with intersections occurring for $CBrF_3$ and $CF_3I$. For SIMPLE gels however, $SF_6$ and $CBrF_3$ at 20°C are both S > 0.7 hence sensitive to complications from spontaneous nucleations and low LET irradiations; $TeF_6$, with melting point -38.9°C and boiling at -37.6°C, is a liquid only in a 1°C window, hence not useful: only $CF_3I$ is S < 0.7.

We examine more closely the cases of $CBrF_3$, $CBrClF_2$ and $CF_3I$ for which complete thermophysical properties are known and $E_c$ can be calculated. The corresponding recoil thresholds of each are shown in Fig. 22, calculated as in Fig. 17. As evident, the results confirm the FMs of Table IV. With $CBrF_3$, an $E_{thr}^{nr} \sim 1$ keV can be achieved at 3ºC and 2 bar (S ~ 0.57); $CBrClF_2$, an $E_{thr}^{nr} \sim 1$ keV at 2 bar and 75ºC (S ~ 0.7). In contrast, $CF_3I$ is only able to provide an $E_{thr}^{nr} \sim 10$ keV at 25ºC (near $T_{gel}$) and 2 bar (S ~ 0.32).

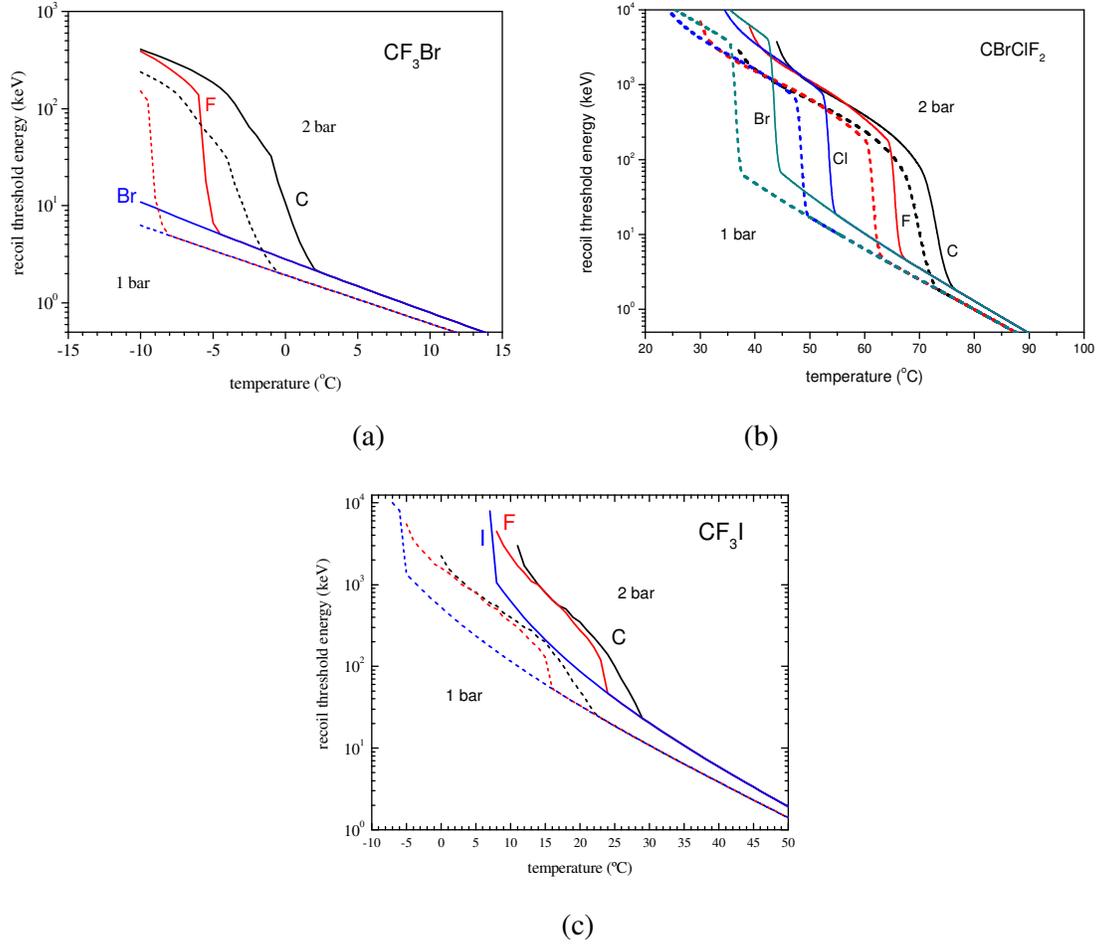

Fig. 22: recoil threshold curves for (a) $CBrF_3$, (b) $CBrClF_2$, and (c) $CF_3I$ with temperature.

## 5.2 Detector Fabrications

The variation of the three liquid densities with temperature are shown in Fig. 23.

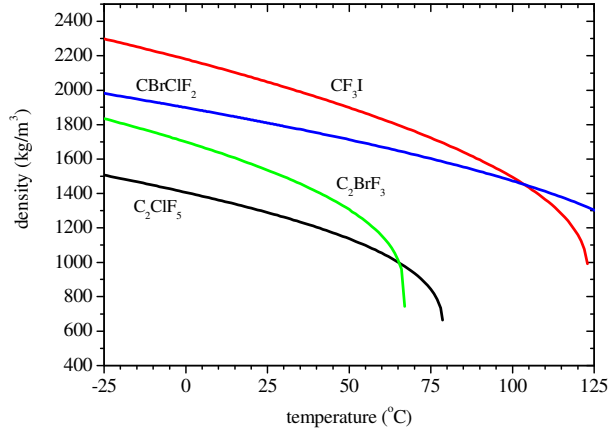

Fig. 23: temperature variation of densities of $CBrF_3$, $CBrClF_2$, and $CF_3I$ relative to $C_2ClF_5$.

As evident from Fig. 23, the significantly higher-density heavy liquid SDD fabrications must generally proceed on the basis of a serious viscosity-matching of the liquid with the gel. An estimate of the minimum viscosity ($\varphi$) required to trap the droplets during the fabrication process is given by [64]

$$\varphi = 2r^2 gt \frac{\rho_l - \rho_g}{9D} \quad , \tag{9}$$

where r is the average droplet radius, D is the height of the gel, t is the time for a droplet to fall a distance D, and $\rho_l$ ($\rho_g$) is the liquid (gel) density. In the case of $CF_3I$, for t = 1 hour (the time required for the setting of the gel during cooling), $\rho_l$ ($\rho_g$) = 2 x $10^3$ kg/m$^3$ (1.3 x $10^3$ kg/m$^3$), r = 35 x $10^{-6}$ m, D = 5 x $10^{-2}$ m, and $\varphi$ = 0.13 kg/m/s.

The gel itself is formed as previously by combining powdered gelatin and bi-distilled water with slow agitation to homogenize the solution; separately, PVP is added to bi-distilled water, and agitated at 60ºC. Pre-eluted ion-exchange resins for actinide removal are added to both, removed by filtering after blending in a detector bottle by agitation. The viscosity variations are effected with a 0.44 wt% agarose addition, effected by combining the additive (Sigma Aldrich A0576) with glycerin at 90ºC to break the agarose chains, and its addition to the concentrated gel mix prior its filtration. Following outgassing and foam aspiration, the solution is left overnight at 42ºC with slow agitation to prevent air bubble formation. The final gel matrix recipe, which produced a uniform and homogeneous distribution of droplets, had a measured $\varphi$ = 0.17 kg/m/s, as well as an increased temperature at which the transition from solution to gel (sol-gel transition) occurs. $CBrF_3$ (ρ ~ 1.5 g/cm$^3$) and $CBrClF_2$ (ρ ~ 1.8 g/cm$^3$) would also require the same fabrication technique, with the advantage of a somewhat smaller agarose addition.

SDD fabrication occurs via the same phase diagram of Fig. 2, adjusted for the pressure and temperature of the liquid. The detector bottle is removed to a hotplate within a hyperbaric chamber, and the pressure raised to just beyond the vapor pressure at 42ºC. After thermalization, the agitation is stopped and the $CF_3I$ storage bottle opened to admit the liquid through the same condensing-distillation line with a 0.2 µm filter used previously.

Once the $CF_3I$ is injected, the pressure is quickly raised to 15 bar to prevent the droplets from rising to the surface, and a rapid agitation initiated to shear big droplets; simultaneously, the temperature was raised to 50ºC to create a temperature gradient inside the matrix and permit dispersion of the droplets. After 20 minutes, the temperature is slightly reduced for 1 hr (with pressure and agitation unchanged). The $CF_3I$, in liquid state, is divided into smaller droplets by the continued agitation. Finally, the heating is stopped: the temperature decreases until the sol-gel transition is crossed, during which the stirring is reduced and finally stopped. The droplet suspension is quickly cooled to 10ºC and left to set for 40 minutes, then cooled to 5ºC where it is maintained for ~ 15 hours. The pressure is then slowly reduced to atmospheric pressure, and the detector removed to cold storage: a fabrication example is shown in Fig. 24.. The process results in approximately uniform and homogeneous (40 ± 15 µm diameter) droplet distributions, as determined by optical microscopy. Longer fractionating times give narrower distributions of smaller diameters; shorter, broader distributions of larger diameters.

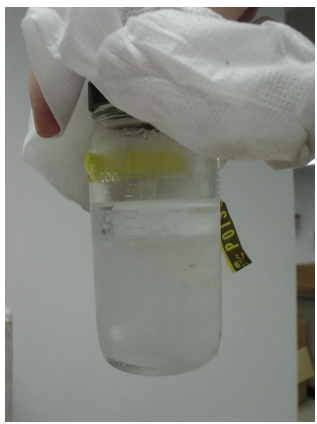

Fig. 24 : Completed $CF_3I$ detector prototype.

**5,3    Solubility and Lifetime**

As stated above, higher density liquids are generally characterized by higher solubilities, which determines the amount of active target mass in the detector, as well as the fracture probability of the gel. Table V indicates the solubilities of the three liquids, all of which are larger than that of $CCl_2F_2$ by a factor of 5-10.

Table V: solubilities of $CBrF_3$, $CBrClF_2$ and $CF_3I$.

| Liquid | Solubility (g/liter$H_2O$/bar at 25°C) |
|---|---|
| $CBrF_3$ | $0.32^{38}$ |
| $CBrClF_2$ | $0.28^{38}$ |
| $CF_3I$ | ~ $0.5^{64}$ |

Unlike previous detectors made with $C_2ClF_5$, the $CF_3I$ prototypes began to significantly fracture within several hours of fabrication. The fracturing is inhibited by overpressuring the devices, but not eliminated. Tests with a SDD made by dissolving the liquid inside the gel produced cracks within 24 hrs, indicating the fracturing to occur because of a high solubility of $CF_3I$ gas inside the gel. Although this phenomenon occurs with or without bubble nucleation, because the $CF_3I$ gas inside the gel occupies any microscopic $N_2$ gas pockets formed during the fractionating stage of the suspension fabrication, it is aggravated by nucleations arising from the ambient background radiations.

Despite the initial fracturing, the $CF_3I$ prototype remained active for almost a year after removal to an underground "cool" storage at 16°C at 2 bar, with little growth of the fractures observed in the measurement [64]. Nevertheless, the problem of fracturing requires an improved understanding of the involved chemistry and development of new techniques, to include the possible use of gelifying agents not requiring water as a solvent or the use of ingredients to inhibit the diffusion of the dissolved gas, which in turn suggests a possible shift to organic gels if the radio-purity of the current gel fabrications can be maintained or improved.

## 5.4 Particle Discrimination

Similar tests made of the CF$_3$I prototype [64] at 35ºC and 1 bar with the instrumentation of the present light experiments under similar experimental conditions yielded signal events with $\mathcal{F}$ = 520±32 Hz, τ = 7.8-21 ms and $\mathcal{A}$ = 160-500 mV, consistent with the light nuclei SDD signals of α origin in Sec. 3.

Irradiations of the small volume device prototypes by $^{60}$Co verified the device insensitivity to γ's below T$_{gel}$, consistent with the general response of SDDs. Irradiations with a filtered neutron beam demonstrated sensitivity to reactor neutron irradiations via the induced recoils of fluorine, carbon and iodine. Fig. 25 displays the results of a 144 keV neutron irradiation of a device at 1 bar, with the rapid rate increase beginning ~ 40ºC consistent with the iodine sensitivity onset observed in the temperature variation of the threshold incident neutron energies. The expected signal from fluorine and carbon at 20ºC is masked by the iodine response to a broad, higher energy neutron component of the filtered beam, as identified in Ref. [45].

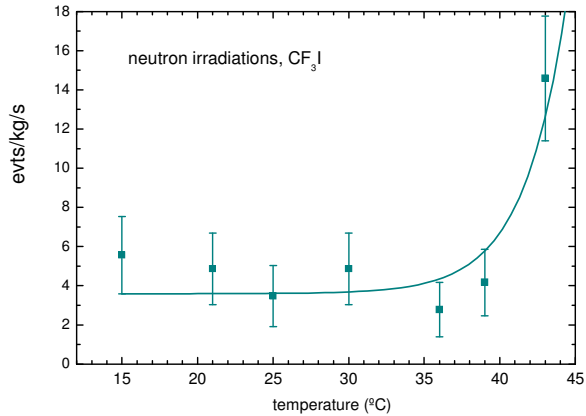

Fig. 25 : 144 keV filtered neutron beam irradiation of a CF$_3$I prototype; the line represents an exponential fit to the data.

## 5.5 Particle Discrimination

With respect to the discussion of particle discrimination in Sec. 4, the critical LET = 76 keV/μm for CF$_3$I at 50ºC and 2 bar: as seen in Fig. 26, although the 5.5 MeV α Bragg peak is shifted to a larger depth, the protobubble production capability ranges 0-47 μm: there is no evident $r_{min}$ in the droplet size, the gap criterion cannot be satisfied, and the resulting $\mathcal{A}$ will likely overlap -- as in fact observed in these measurements which yielded 4 events with $\mathcal{A}_\alpha < 100$. For $E_\alpha = 8.0$ MeV, a $r_{min} \sim 14$ μm exists, but remains unlikely to provide the gap. At 1 bar operation, the critical LET = 63 keV/μm, there is again no $r_{min}$ and no simple discrimination seems possible.

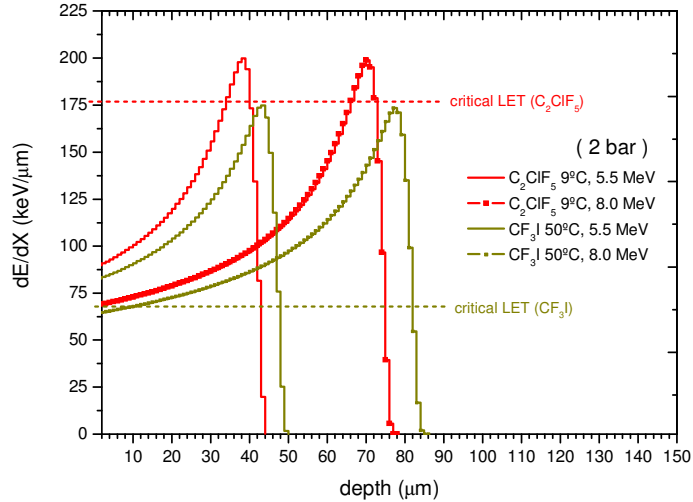

Fig. 26: comparison of the energy loss depth profiles for 5.5 and 8.0 MeV α's in $CF_3I$ and $C_2ClF_5$ at 2 bar and representative low $E_{thr}$ temperatures.

Thus it would appear that in dark matter search applications, a SIMPLE $CF_3I$ device would be unable to provide a complete particle discrimination for the U/Th α's without resorting to FFT integrations as employed by COUPP. Again however, as with $C_4F_{10}$ we stress that the critical LET is dependent on $\Lambda$ which is also not well-known for these heavier liquids; for $CF_3I$ however, use of $\Lambda = 4$ as in Ref. [2] would lower the critical LET, exacerbating the situation.

## *6.* **Conclusions**

SDDs with the light and heavy nuclei liquids in this study can be fabricated with the SIMPLE food-based gel, via either density- or viscosity-matching using appropriate protocols and gel chemistry to provide a homogeneous, reproducible, well-defined distribution of droplet sizes. The result is detectors with approximately the same response capability – although the operational temperatures and pressures to achieve a given $E_{thr}^{nr}$ are necessarily different, and constrained by the proximity of the SDD operating conditions to the melting point of the gel as well as the liquid solubility.

In contrast to PICASSO and COUPP, the characteristics of all particle-generated events of the various SDDs lie within the ranges previously defined for the $C_2ClF_5$ device with α-generated events, which we suspicion is attributable in part to the gel presence/nature -- but further study is required to confirm.

The signal response of the SDD in the case of particle-induced events is largely dependent on the droplet size distribution, which depends on the fractionating speed and time, and can be varied to yield differing distributions. For dark matter searches, discrimination between α and nuclear recoil events appears to depend on the relation between the droplet size distribution (which determines the recoil event spectrum), the background α Bragg peak in the liquid and its component ≥ critical LET, with the indication that neither $C_4F_{10}$ or $CF_3I$ in a SIMPLE configuration is able to provide a clear particle discrimination. Given however the lack of a complete understanding of the observed gap formation and liquid Λ, further research is required and the particle discrimination capacity of each SDD must at present be determined experimentally.

Thus said, the choice of SDD liquid remains fundamentally dependent on the required operating conditions to achieve both low $E_{thr}^{nr}$ and particle discrimination. Of the light nuclei liquids, PICASSO, using $C_4F_{10}$ operated at 50-60ºC and 1 bar, obtains a $E_{thr}^{nr}$ ~ 1.7 keV for neutron-generated recoils, but without well-defined particle-discrimination. SIMPLE, using $C_2ClF_5$ with its food gel, runs at 9ºC and 2 bar for a recoil $E_{thr}^{nr}$ = 8 keV, with an operating range generally limited to < 15ºC because of the onset of Cl sensitivity to γ's; for $E_{thr}^{nr}$ ≤ 8 keV, neither $C_4F_8$ or $C_4F_{10}$ seems usable in a SIMPLE device for WIMP search applications, given their $E_{thr}^{nr}$ at $T_{gel}$. Use of a different gel (as in early PICASSO) is possible, but the questions of increased backgrounds and particle discrimination would need to be addressed (possibly, using the PICASSO and COUPP analyses techniques).

The light nuclei devices described here, while suffering from the $A^2$ enhancement of the heavy liquids in the SI sector, are still capable of contributing to this sector if they can be operated at temperatures and pressures corresponding to $E_{thr}^{nr}$ ~ 2 keV, as in the recent case of PICASSO, owing to the flattening of the exclusion curves with decreasing $E_{thr}^{nr}$. The liquid selection for SIMPLE devices is however constrained by its gel nature to $C_2ClF_5$, $C_3F_8$ and $CBrF_3$ because – all else being equal – of their ability to achieve $E_{thr}^{nr}$ < 4 keV at temperatures < $T_{gel}$. Of these, $C_3F_8$ provides the lowest $E_{thr}^{nr}$: a simultaneous measurement with separate SDDs of $C_3F_8$ and $CF_3Br$, operated at 15ºC and 1 bar, could theoretically provide $E_{thr}^{nr} \sim 3$ keV in both cases.

Numerous heavier target liquid possibilities exist which would provide, assuming SDD fabrication feasibility based on viscosity-matching or development of more temperature-resilient gels such as PICASSO's earlier polyacrylamide, an increased sensitivity in the SI sector as well as both sectors of the SD studies. Introduction of FM = T (S = 0.7) permits a pre-selection among the possibilities in terms of dark matter search suitability. Further investigations of their liquid phase parameters (as well as commercial availability, price and environmental impact) is however required before decisions can be taken in their regard, as also the development of new gels capable of supporting the thermodynamic conditions necessary to a low $E_{thr}^{nr}$ operation and particle discrimination.

## Acknowledgements

We thank A.R. Costa for assistance in the production of the SDDs, and M. Silva for the construction of the hermetic device caps. The activity of M. Felizardo was supported by grant SFRH/BD/46545/2008 of the Portuguese Foundation for Science and Technology (FCT). The activity was supported in part by POCI grant FP/63407/2005 of FCT, co-financed by FEDER, by FCT POCTI grant FIS/55930/2004, and by FCT PTDC grants FIS/115733/2009 and FIS/121130/2010.